\documentclass[epj-spec]{svjour}
\usepackage{graphicx,float}
\usepackage{amsmath,amssymb}
\usepackage{color}

\newif\ifhires \hiresfalse
\newcommand{\w}{\omega}
\newcommand{\TK}{T_{\rm K}}

\newcommand{\rmax}{r_{\rm max}}
\newcommand{\Simp}{S_{\rm imp}}

\newcommand{\deff}{d_{\rm eff}}

\input{tcilatex}

\begin{document}

\title{ Critical quasiparticles in single-impurity and lattice Kondo models }
\author{Matthias Vojta\inst{1} \and Ralf Bulla\inst{2} \and Peter W\"olfle\inst{3}}
\institute{Institut f\"ur Theoretische Physik, Technische Universit\"at Dresden, 01062
Dresden, Germany \and
Institut f\"ur Theoretische Physik, Universit\"at zu K\"oln, Z\"ulpicher Stra{\ss }e 77, 50937 K\"oln, Germany \and
Institut f\"ur Theorie der Kondensierten Materie, Karlsruher Institut f\"ur Technologie, 76049 Karlsruhe, Germany}


\abstract{
Quantum criticality in systems of local moments interacting with itinerant electrons has become an important and diverse field of research. Here we review recent results which concern (a) quantum phase transitions in single-impurity Kondo and Anderson models and (b) quantum phase transitions in heavy-fermion lattice models which involve critical quasiparticles. For (a) the focus will be on impurity models with a pseudogapped host density of states and their applications, e.g., in graphene and other Dirac materials, while (b) is devoted to strong-coupling behavior near antiferromagnetic quantum phase transitions, with potential applications in a variety of heavy-fermion metals.
}

\maketitle


\section{Introduction}

\label{intro}

Quantum phase transitions (QPT) \cite{ssbook,mvrop,lrvw07,ss_rev08,steg_rev08,gia_rev08} continue to be an exciting field of research in condensed matter physics, both theoretically and experimentally. While for the simplest situations, such as magnetic transitions in Mott insulators devoid of quenched disorder, agreement between theory and experiment has
been achieved, more complicated cases are far from being fully understood.
These include
(i) quantum phase transitions with conventional, i.e., symmetry-breaking, order parameters in \emph{metallic} systems \cite{lrvw07},
(ii) quantum phase transitions under the influence of quenched disorder \cite{tv_rev06},
(iii) interaction-driven metal-to-insulator transitions \cite{imada_rev98},
(iv) topological transitions in interacting systems \cite{xu12,fiete12},
and (v) boundary quantum phase transitions \cite{mvrev}.
Each of these cases poses its specific challenges to theory, and the lack of a complete theoretical framework prohibits an efficient classification and understanding of experimental observations.

In this paper, we will provide a partial review on strands of work performed in the context of Kondo systems \cite{hewson} which belong to the cases (i) and (v), namely antiferromagnetic transitions in strongly interacting metals and transitions in variants of the single-impurity Kondo model, describing
isolated magnetic moments immersed in baths of fermionic quasiparticles. The conceptual link is given by the occurrence of ``critical'' fermionic quasiparticles, where fermionic spectral functions display power-law singularities. In addition, we will highlight connections to case (iii), namely to Mott transitions and partial (or orbital-selective) Mott transitions where critical quasiparticles are expected as well.

The body of the paper is organized as follows:
In Section~\ref{sec:cqp} we will discuss conceptual aspects of critical quasiparticles and provide a
quick survey on available results in the literature.
Section~\ref{sec:pg} is devoted to the Kondo model with a power-law density of states. We summarize the theoretical understanding of its quantum phase transitions and discuss some recent applications, including Kondo impurities in graphene and on the surface of topological insulators.
Section~\ref{sec:pw} reviews a recently developed theory for antiferromagnetic bulk quantum phase transitions, which assumes strong coupling between fermions and order-parameter fluctuations such that power-law singularities develop in the fermionic spectrum on the entire Fermi surface.
A summary and outlook will be given in Section~\ref{sec:summ}.


\section{Critical quasiparticles}
\label{sec:cqp}

Phase transitions into symmetry-breaking phases are characterized by the critical behavior of a collective order parameter \cite{ssbook}. This order parameter is a bosonic field; for transitions of electrons in solids it can usually be represented as a composite of two (or four) fermion operators. In contrast, single-particle excitations are not generically critical at the transition point. However, recent research has highlighted the possibility -- and potential experimental relevance -- of critical single-particle excitations at certain quantum phase transitions -- these form the subject of this review.

In both lattice and continuum systems, power-law behavior of fermionic single-particle propagators may occur as function of frequency or wavevector, or both. As fermionic low-energy excitations of stable metals occur on Fermi surfaces (more generally hypersurfaces) in momentum space, critical power-law singularities may similarly occur on surfaces in momentum space: Most naturally, the Fermi surface of a metal may develop into a surface of critical quasiparticles upon approaching the quantum critical point.\footnote{We use the term ``critical'' in the context of fermionic spectral functions if the quasiparticle weight vanishes, i.e., the $\delta$ peak is replaced by a critical power-law continuum as function of frequency. We note that such power-law spectral functions are realized, e.g., in one-dimensional interacting electron systems described by Luttinger-liquid theory.} This is distinct from the standard bosonic order-parameter situation, where singular behavior of correlation functions at criticality is usually restricted to the vicinity of a single point in energy--momentum space.

Critical fermionic quasiparticles have been discussed in various contexts.
On the one hand, they may occur as ``active'' critical degrees of freedom at Mott metal--insulator transitions: This has been proposed within effective theories for Mott and orbital-selective Mott transitions \cite{flst2,senthil08,osmottrev}. For such transitions there is no local order parameter, but the loss of metallicity accompanied by a change in the Fermi volume is a defining criterion. Available analytical theories have employed representations in terms of slave particles and gauge fields; in this language the Mott transition is often marked by the condensation of an auxiliary boson coupled to a gauge field.
Recent numerical findings \cite{dobro13} of quantum critical scaling of the conductivity near the Mott critical endpoint as described by dynamical mean-field theory also hint at fermionic criticality.

On the other hand, fermions of a metal may couple to critical order-parameter degrees of freedom in such a way that quasiparticles are rendered critical by this coupling \cite{lrvw07}. In general, this may happen in a momentum-selective fashion, such that, e.g., only fermions at so-called hotspots (momenta connected by the wavevector of the order parameter) become critical. More interesting are situations where fermions at all (or almost all) Fermi-surface momenta become critical.
The latter applies to the particular strong-coupling theory of antiferromagnetic quantum phase transitions which will be reviewed in Section~\ref{sec:pw}. It applies as well to Ising-nematic quantum phase transitions in 2d \cite{metzner03} where the ordering wavevector is zero and hence all momenta are hotspots. Finally, singular behavior of fermions at all momenta also underlies the idea of local quantum criticality in the framework of extended dynamical mean-field theory \cite{edmft}.

For quantum phase transitions in fermionic quantum impurity models \cite{mvrev}, power-law behavior of {\em local} propagators is not uncommon, and we will review a well-studied case in Section~\ref{sec:pg}. In some of the models, fermions can indeed be interpreted as the critical degrees of freedom. An understanding of these types of impurity criticality (being often more tractable than lattice cases) may also be relevant to the critical behavior of lattice models as described by dynamical mean-field theory \cite{dmft} and extensions thereof.


\section{The pseudogap Kondo effect}
\label{sec:pg}

As a well-understood example for critical quasiparticles, we discuss the rich physics of the so-called pseudogap Kondo model. This model has appeared first in the context of magnetic impurities in unconventional superconductors \cite{withoff}. While this initial work only uncovered the existence of a quantum phase transition between a screened and an unscreened impurity moment, detailed numerical studies \cite{ingersent,bulla97,GBI,si02} using Wilson's numerical renormalization group (NRG) technique \cite{nrg,nrg_rev} determined the complete phase diagram and some of the critical properties. A complete analytical understanding of the low-energy physics was achieved \cite{VF04,FV04} after realizing that the Anderson model provides a framework which allows to utilize controlled epsilon-expansion techniques. Later, the pseudogap Kondo model was applied to describe impurities in graphene and other Dirac materials \cite{fvrop}.

In the following we review the basic aspects of the pseudogap Kondo problem, with an eye on critical quasiparticles, and discuss recent applications. We will restrict our attention to the case of a spin $S=1/2$ coupled to a single screening channel; the two-channel version has been discussed in detail in Refs.~\cite{GBI,schneider11}.

\subsection{The pseudogap Kondo and Anderson models}
\label{sec:pgmodel}

The standard Kondo Hamiltonian~\cite{hewson} reads
\begin{eqnarray}  \label{eq:Kondomodel}
\mathcal{H} = \sum_{{\vec{k}},\sigma} \epsilon_{{\vec{k}}} c^\dagger_{\vec{k}%
\sigma} c^{\phantom{\dagger}}_{\vec{k}\sigma} +V_0\sum_{{\vec{k}},{\vec{k}}%
^{\prime },\sigma} c^\dagger_{\vec{k}\sigma} c^{\phantom{\dagger}}_{\vec{k}%
^{\prime }\sigma} +J_0\; {\vec{S}} \cdot {\vec{s}}_0,
\end{eqnarray}
where $\vec{S}$ is the impurity spin $S=1/2$, $c_{\vec{k}\sigma}$ are conduction-electron operators, and ${\vec{s}}_0=\frac{1}{2}\sum_{{\vec{k}}{\vec{k}}^{\prime }} c^{\dagger}_{\vec{k}\sigma} {\vec{\tau}}_{\sigma \sigma^{\prime }}c^{\phantom{\dagger}}_{\vec{k}^{\prime }\sigma^{\prime }}$ is their spin density at the impurity site, with ${\vec{\tau}}$ the vector of Pauli matrices. The Kondo coupling $J_0$ and the potential-scattering strength $V_0$ characterize the impurity. In the following, we denote the density of states (DOS) of the conduction band at the impurity site by $\rho(\w)$.

The Kondo model may be derived from the more general Anderson model,
\begin{equation}\label{eq:AIM}
\mathcal{H} = \sum_{{\vec{k}},\sigma} \epsilon_{{\vec{k}}} c^\dagger_{{\vec{k}}\sigma}
c^{\phantom{\dagger}}_{{\vec{k}}\sigma} +\epsilon_d \sum_\sigma n_{d\sigma}
+ U n_{d\uparrow} n_{d\downarrow} + \sum_{\vec{k},\sigma} \left( v_{\vec{k}} c^\dagger_{{\vec{k}}\sigma} d^{\phantom{\dagger}}_\sigma + h.c.
\right)
\end{equation}
which describes an impurity level $d$, with $n_{d\sigma} =  d^\dagger_\sigma d^{\phantom{\dagger}}_\sigma$, hybridized with a conduction band. In this model, the $d$ level prefers single occupancy if $\epsilon_d<0$ and $\epsilon_d+U>0$. For large $|\epsilon_d|$ and $U$, charge fluctuations are frozen out, such that an effective spin 1/2 degree of freedom remains whose dynamics can be mapped to a Kondo model \eqref{eq:Kondomodel}. Second-order perturbation theory yields \cite{hewson}:
\begin{equation}
J_0 = 2 v^2 \bigg(\frac{1}{|\epsilon_d|} + \frac{1}{|U+\epsilon_d|}\bigg),~~
V_0 = \frac{v^2}{2} \bigg(\frac{1}{|\epsilon_d|} - \frac{1}{|U+\epsilon_d|}%
\bigg)  \label{swo}
\end{equation}
where $v_{\vec{k}}\equiv v$ has been assumed.

For a metallic host, the DOS $\rho(\omega)$ is finite at the Fermi level. Then, for antiferromagnetic $J_0>0$ the impurity spin is screened below the so-called Kondo temperature $\TK$. For a flat conduction-band DOS, $\rho(\omega)=\rho_0$, one finds~\cite{hewson}:
\begin{eqnarray}  \label{tkeq}
\TK = \sqrt{D J_0}\, e^{-1/(J_0 \rho_0)}\;.
\end{eqnarray}

In the following, we will instead concentrate on the case of Kondo impurities coupled to fermions with a pseudogap DOS, of the low-energy form $\rho(\omega) \propto |\w|^r$. For $r>0$ this DOS corresponds to a semimetal; it can be thought of arising from linearly dispersing Dirac electrons in $(1+r)$ space dimensions, with $r=1$ corresponding to graphene.
As a consequence of the vanishing DOS at the Fermi level the tendency toward Kondo screening is reduced, such that no screening occurs at small Kondo coupling $J_0$, and a quantum phase transition between phases without and with screening occurs upon increasing $J_0$ \cite{mvrev,withoff}.
As the critical behavior discussed below is independent on high-energy details of the DOS, we parameterize $\rho(\w)$ as
\begin{equation}
\rho(\omega) = \frac{1+r}{2D^{r+1}}\,|\omega|^r\,\Theta(|\omega|-D)
\label{pgdos}
\end{equation}
with $D$ being a bandwidth.

\subsection{Phase diagram}

\label{sec:pd}

Despite its simplicity, the pseudogap Kondo model has an extraordinarily rich phase diagram, first determined by Gonzalez-Buxton and Ingersent \cite{GBI} using NRG. The physics depends not only on $J_0$ and the exponent $r$ of the low-energy DOS, but also on the presence or absence of particle--hole (p-h) symmetry. Importantly, p-h symmetry requires \emph{both} $\rho(\omega) = \rho(-\omega)$ in the host and $V_0=0$ in the Kondo model \eqref{eq:Kondomodel} [or $U=-2\epsilon_d$ in the Anderson model \eqref{eq:AIM}].

We start by summarizing the phase diagram of the pseudogap Kondo model. We restrict our attention to the case of antiferromagnetic Kondo coupling, $J_0>0$, and use the acronyms of Ref.~\cite{GBI} for both phases and fixed points.

\begin{figure}[t]
\center
\includegraphics[width=0.9\textwidth]{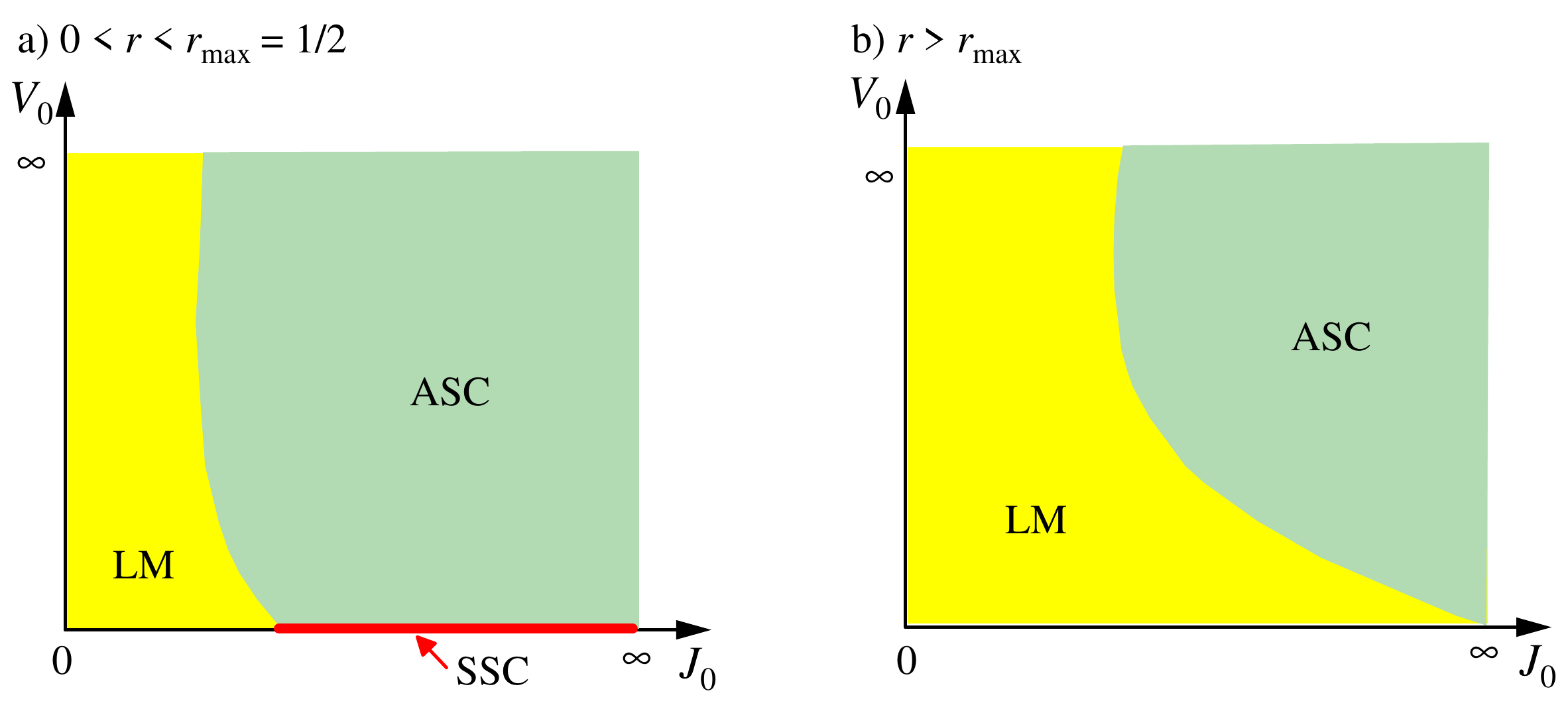}
\caption{ Schematic phase diagram for the pseudogap Kondo model in the plane spanned by the Kondo coupling $J_0$ and the potential scattering $V_0$, the latter measuring particle--hole asymmetry. The phases are denoted as local-moment phase (LM), symmetric strong-coupling phase (SSC), and asymmetric strong-coupling phase (ASC).
For $0<r<\rmax=1/2$ screening can occur at particle--hole symmetry (left), whereas screening is only possible in the presence of particle--hole asymmetry for $r>\rmax$ (right).
In addition to the change at $\rmax$, the structure and nature of the critical fixed points also changes at $r^\ast = 0.375\pm0.002$ and $r=1$, for details see text.
}
\label{fig:pd}
\end{figure}

The phase diagram, qualitatively sketched in Fig.~\ref{fig:pd}, shows that Kondo screening is only realized for large $J_0$; in addition p-h asymmetry is favorable for screening. There are three stable phases: the local-moment phase (LM), the p-h symmetric strong-coupling phase (SSC), and the p-h asymmetric strong-coupling phase (ASC). Here, LM corresponds to an asymptotically free (unscreened) spin, with a residual entropy $\Simp=\ln2$. SSC represents the generalization of the metallic Kondo-screened phase to finite $r$. Due to the vanishing host DOS, SSC is characterized by a residual entropy $\Simp=2r\ln 2$, implying {\em partial} screening. Moreover, SSC only exists for $r<1/2$. Finally, ASC is a fully screened phase with vanishing residual entropy. In stark contrast to the metallic case $r=0$ where p-h asymmetry is marginally irrelevant, we see that p-h asymmetry is relevant for $r>0$, i.e., SSC is destabilized by any amount of p-h asymmetry.

As deduced from the numerical solution of the pseudogap Kondo model \cite{GBI}, the topology of the phase diagram changes \emph{qualitatively} as the bath exponent $r$ is varied, see Fig.~\ref{fig:pd}.

\paragraph{a) $\mathbf{0<r< \rmax=1/2}$.}

For particle--hole symmetry, a critical coupling $J_{\mathrm{c}}$ separates LM from SSC: For initial values $J<J_{\mathrm{c}}$ the system is in the LM phase, whereas it is in the SSC phase for $J>J_{\mathrm{c}}$.
For finite particle--hole asymmetry, i.e. $V_0\neq 0$, there is a boundary between LM and ASC.

\paragraph{b) $\mathbf{r>\rmax}$.}

SSC disappears such that there is no Kondo screening at particle--hole symmetry, irrespective of the strength of the Kondo coupling $J_0$.
In contrast, screening is still possible for finite asymmetry, where a transition between LM and ASC continues to exist.

\paragraph{c) $\mathbf{-1<r<0}$.}

This regime can possibly be realized in the case of reconstructed vacancies in graphene~\cite{cazalilla12,mf13} but was analyzed more generally in Ref.~\cite{fracsp}. SSC is stable, and a phase transition separates SSC from a newly emerging phase ALM, corresponding to $J_0=0$ and large particle--hole asymmetry. In the following we will, however, not discuss $r<0$ in any detail.

\subsection{Quantum phase transitions}

The pseudogap Kondo model features two qualitatively different transitions: The transition between LM and SSC is controlled by a p-h symmetric critical (SCR) fixed point, while the transition between LM and ASC is controlled by a p-h asymmetric critical (ACR) fixed point (for $r>r^\ast$, see below). The existence of SCR is tied to that of the SSC phase; SCR is thus present for $0<r<1/2$.

Interestingly, also ACR exists in a restricted $r$ range only, namely for $r>r^\ast\approx0.375$. As $r$ approaches $r^\ast$ from above, ACR merges with SCR, such that for $0<r<r^\ast$ the transition between LM and ASC is controlled by SCR which is then a multicritical point. Further, the critical exponents of ACR become trivial for $r>1$, showing that $r=1$ plays the role of an upper critical dimension \cite{VF04}.

The pseudogap Kondo and Anderson models share identical fixed points and quantum phase transitions \cite{GBI}. This observation can be rationalized within the effective field theories described in what follows.


\subsection{Critical field theories}
\label{sec:field}

The complicated topology of the RG flow \cite{GBI} suggests that different field theories are required to describe the critical properties near the SCR and ACR fixed points. Such field theories have been worked out in detail in Refs.~\cite{VF04,FV04} and provide an essentially complete analytical understanding of the pseudogap Anderson and Kondo models. Interestingly, none of these field theories is of conventional (i.e. bosonic) Landau-Ginzburg-Wilson type; instead all are of genuinely fermionic character and are formulated in the degrees of freedom of either the Kondo or the Anderson model.

In the following we shall summarize the three relevant critical theories. When specifying flow equations from perturbative RG, we will assume a symmetric pseudogap density of states as in Eq.~\eqref{pgdos}. The effect of a high-energy particle--hole asymmetry in the DOS can be absorbed in
the impurity part of the Hamiltonian, e.g., the potential scattering term of the Kondo model. This can be rationalized within RG, where integrating out the particle--hole asymmetric piece of the bath at high energies yields an effective model with particle-hole symmetric bath at low energies and a
renormalized impurity Hamiltonian, where in particular the particle--hole asymmetry is accumulated.

\subsubsection{SCR: Kondo model}

For small $r$, an efficient description of the physics at SCR is obtained via the Kondo model itself, Eq.~\eqref{eq:Kondomodel}. A perturbative expansion can be performed in $J_0$ and $V_0$ around the LM fixed point where $J_0=V_0=0$ \cite{withoff,si02,FV04} -- this can be understood as a generalization of Anderson poor man's scaling \cite{poor} to the pseudogap case. As is standard practice, we introduce dimensionless couplings $j$ and $v$, for details see Ref.~\cite{FV04}. Power counting reveals that both couplings are marginal for $r=0$ and irrelevant for $r>0$, $\dim[j] = \dim[v] = -r$. The one-loop flow
equations read
\begin{eqnarray}
\frac{dj}{d \ln D}&=&r j -j^2 \quad \mathrm{and} \quad \frac{dv}{d \ln D}=r
v \;,  \label{poorflow}
\end{eqnarray}
where $D$ denotes the running UV cutoff, initially set by the width of the host band. Eq.~\eqref{poorflow} yields a critical fixed point (SCR) at $j^\ast=r+\mathcal{O}(r^2)$, $v^\ast=0$, which separates the flows towards weak and strong coupling. Controlled calculations near SCR are therefore possible in a double expansion in $r$ and $j$. Potential scattering is irrelevant at SCR and consequently does not play a role for leading critical exponents.

Comparing these properties with the numerically determined phase diagrams one immediately concludes that this critical theory only applies to $0<r<\rmax$, as otherwise no transition occurs at particle--hole symmetry. Indeed, a perturbative calculation of static critical properties of SCR using the Kondo expansion shows excellent agreement with NRG results for small $r \lesssim 0.2$ \cite{FV04}.

\subsubsection{SCR: Symmetric Anderson model}

The Anderson impurity model turns out to provide the relevant degrees of freedom to describe pseudogap Kondo criticality for all $r>0$ \cite{FV04}. To discuss the critical behaviour near SCR, we consider a symmetric Anderson model, Eq.~\eqref{eq:AIM} with $\epsilon_d = -U/2$, a momentun-independent hybridization $v$, and a particle--hole symmetric bath DOS as in Eq.~\eqref{pgdos}. The point $\epsilon_d=U=v=0$ is referred to as the free-impurity fixed point (FImp), whereas the parameter sets $v=0$ and $\epsilon_d = -U/2= \pm\infty$ correspond to doubly degenerate local-moment states in the charge and spin channel, respectively. Therefore $v=0$, $\epsilon_d=-\infty$ can be identified with the LM fixed point, while $v=0$, $\epsilon_d=\infty$ is dubbed LM'.

Notably, the Anderson model is exactly solvable for any $v$ at $U = 0$, known as resonant-level model. In the particle--hole symmetric case, its low-energy physics can be identified with that of the SSC fixed point introduced above: its properties correspond to a partial screening of the impurity degrees of freedom, with a residual entropy of $\Simp = 2r \ln 2$ \cite{GBI,FV04}.

\begin{figure}[t]
\includegraphics[width=0.33\textwidth]{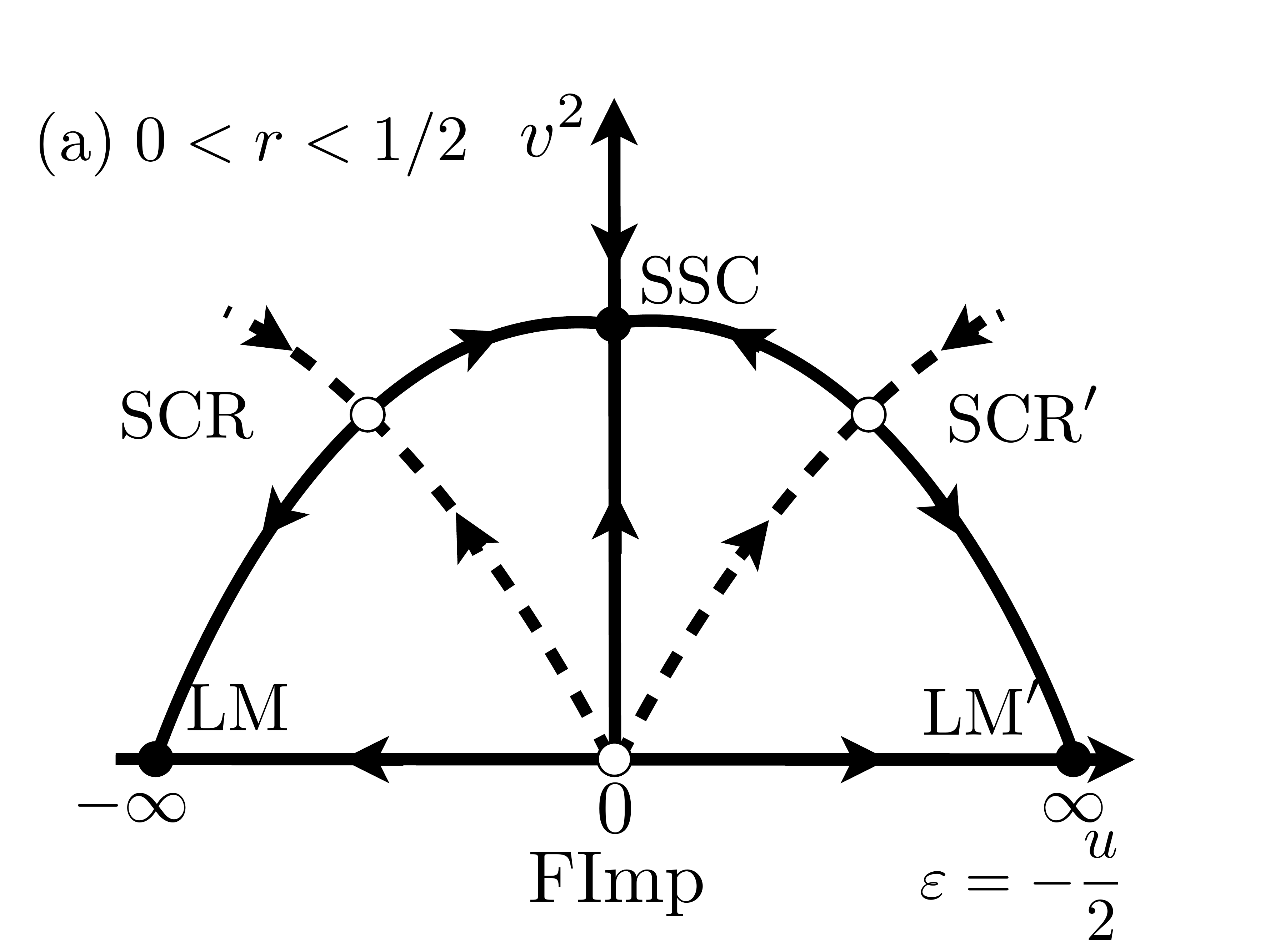} \includegraphics[width=0.33%
\textwidth]{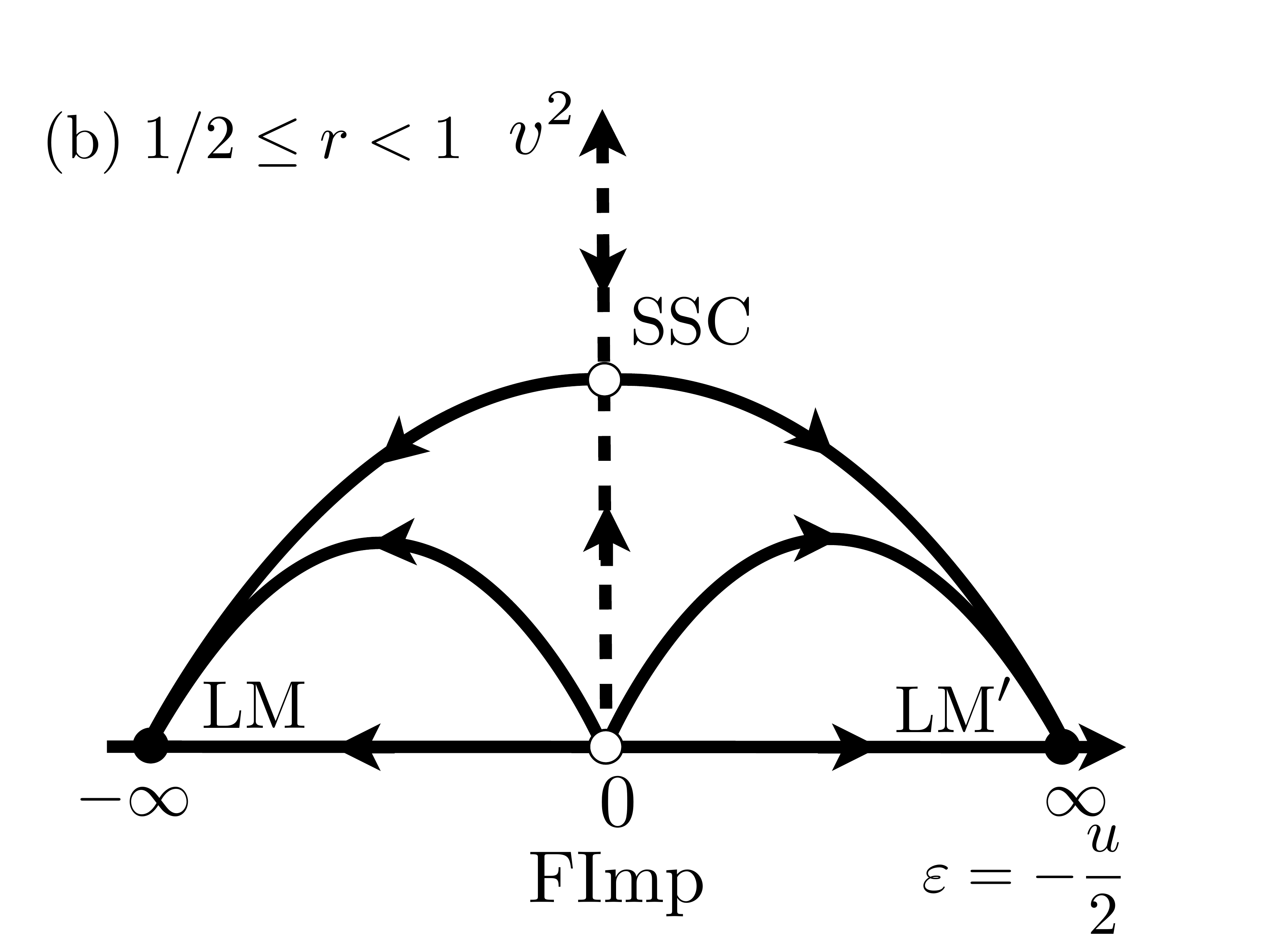} \includegraphics[width=0.33\textwidth]{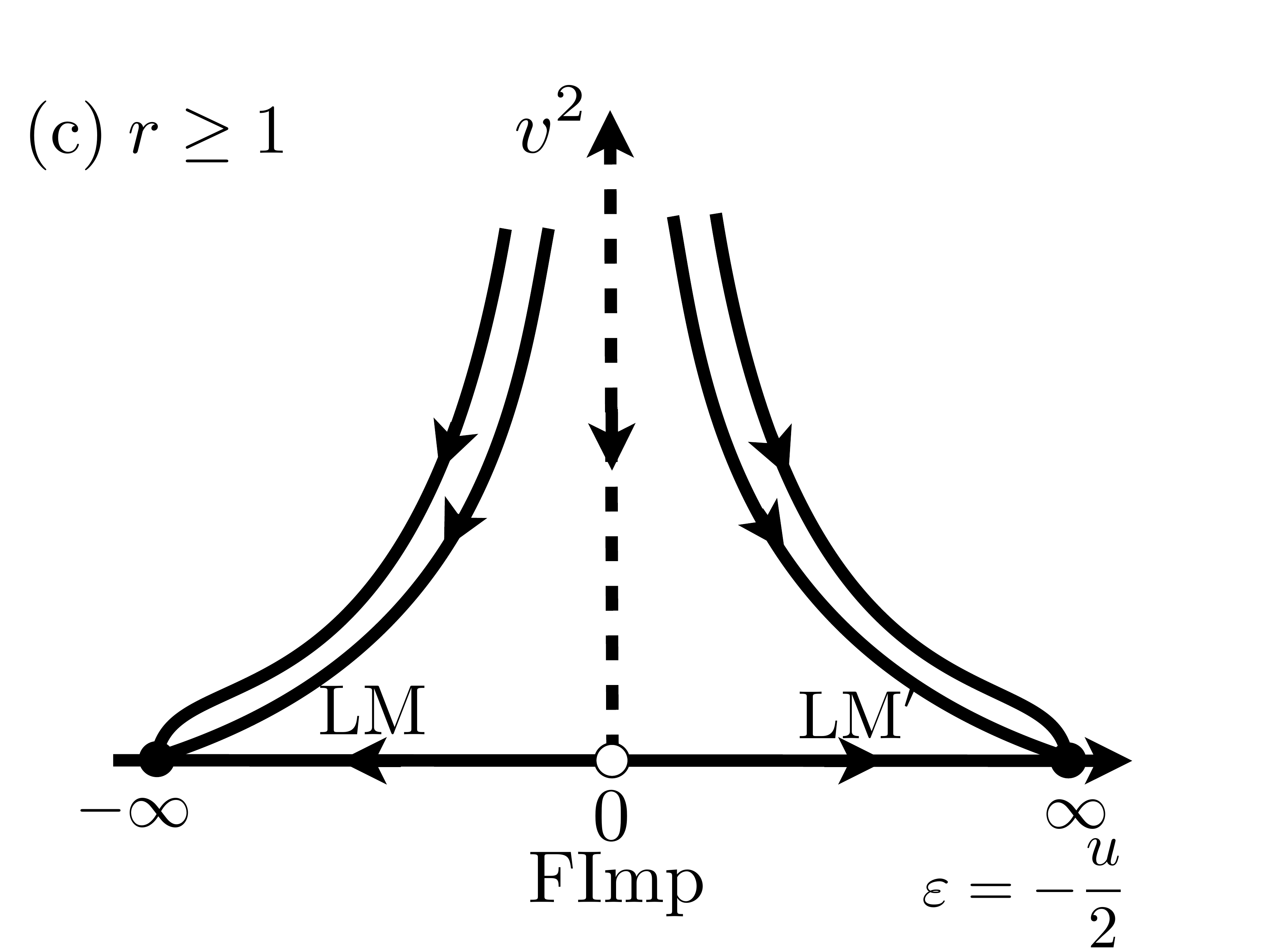}
\caption{
RG flow diagrams for the particle--hole symmetric Anderson model \protect\cite{FV04}, in the plane spanned by the level energy $\epsilon=-u/2$ and the hybridization $v^2$. The symbols for stable fixed points correspond to those for the phases in Fig.~\protect\ref{fig:pd}, with LM and LM' corresponding to local moments formed in the spin or charge channel; the flow of $u$ near SSC is in Eq.~\eqref{eq:rgsym}.
(a) $0<r<1/2$: The critical fixed point SCR (SCR') divides
the flow to LM (LM') from that to SSC. (b) $1/2 \leq r <1$: SCR and SCR'
merge with SSC as $r\to1/2^-$, such that SSC is now unstable. (c) $r\geq1$:
SSC merges with FImp at $\protect\epsilon=v=0$ as $r\to 1^-$. For all $%
r\geq1/2$, LM and LM' are the only stable phases in the presence of
particle--hole symmetry. Reproduced from Ref.~\cite{fvrop}.
}
\label{fig:flowsymmetric}
\end{figure}

A perturbative expansion is now possible in $U$ around the SSC fixed point.
The scaling dimension of the renormalized Coulomb interaction $u$ at SSC is
found to be $\dim[u] = -\overline{r}=-(1-2r)$. The RG flow of $u$ to
two-loop order reads \cite{FV04}
\begin{eqnarray}  \label{eq:rgsym}
\frac{du}{d\ln D}=\left(1-2r\right)u-\frac{3\left(\pi-2 \ln 4\right)}{\pi^2}%
u^3 \;.
\end{eqnarray}
This flow, together with the trivial flow near LM, LM', and FImp is
illustrated in Fig.~\ref{fig:flowsymmetric}. For all $r>0$, LM
is a stable fixed point, while SSC is stable only for $r<\rmax$,
as can be seen from Eq.~\eqref{eq:rgsym}. Therefore, a critical fixed point
(SCR) emerges for $0<r<\rmax$, Fig.~\ref{fig:flowsymmetric}b,
consistent with Fig.~\ref{fig:pd}. Its properties can now be accessed
in a double expansion in $\overline{r}$ and $u$, and Eq.~\eqref{eq:rgsym} yields for the fixed-point coupling at ${u^*}^2=\frac{\pi^2}{3\left(\pi-2\ln 4\right)} \overline{r}$. A perturbative calculation of static critical
properties again yields excellent agreement with NRG results, here for $r\lesssim \rmax$ \cite{FV04}.

Owing to particle--hole symmetry, the behavior at $\epsilon\geq 0$ is formally identical to that at $\epsilon\leq 0$, Fig.~\ref{fig:flowsymmetric}, with the latter describing spin-Kondo physics while the former corresponds to charge-Kondo physics.
SCR' is a critical fixed point between a (partially) screened impurity at SSC and
one with an unscreened charge degree of freedom at LM'.

\subsubsection{ACR: Asymmetric Anderson model.}

We now turn to the ACR fixed point present for $r>r^\ast$. It was realized in Ref.~\cite{VF04} that the critical theory for ACR is that of a level crossing of a many-body singlet and a many-body doublet, minimally coupled to conduction electrons. Using the notation of Ref.~\cite{VF04,FV04}, its Hamiltonian can be written as
\begin{eqnarray}  \label{aiminfu}
\mathcal{H} = \sum_{{\vec{k}},\sigma} \epsilon_{{\vec{k}}} c^\dagger_{{\vec{k}}\sigma} c^{\phantom{\dagger}}_{{\vec{k}}\sigma} + \varepsilon_0
|\sigma\rangle\langle \sigma| + g_0 \left[|\sigma\rangle\langle s|
c_\sigma(0) + \mathrm{h.c.}\right]
\end{eqnarray}
where $|\sigma\rangle=|\uparrow\rangle,|\downarrow\rangle$ and $|s\rangle$
represent the three allowed impurity states. $\varepsilon_0$ is the tuning
parameter (``mass'') of the QPT, i.e. the (bare) energy difference between
doublet and singlet states. The QPT occurs at some $\varepsilon_0=%
\varepsilon_c$, with screening present for $\varepsilon_0>\varepsilon_c$.
Remarkably, this theory is identical to a maximally particle--hole
asymmetric Anderson impurity model, Eq.~\eqref{eq:AIM}, where the doubly
occupied state has been projected out, $U\to \infty$, and $%
(\varepsilon_0,g_0)$ in Eq.~\eqref{aiminfu} have been identified with $%
(\epsilon_d,v)$ in Eq.~\eqref{eq:AIM}.

In this model, the point $\varepsilon_0 = g_0 = 0$ is dubbed valence-fluctuation fixed point (VFl). As above, $g_0=0$, $\varepsilon_0=-\infty$ corresponds to LM, while $g_0=0$, $\varepsilon_0=\infty$ describes a fully screened and particle--hole asymmetric singlet state, to be identified with ASC.

\begin{figure}[tbp]
\includegraphics[width=0.41\textwidth]{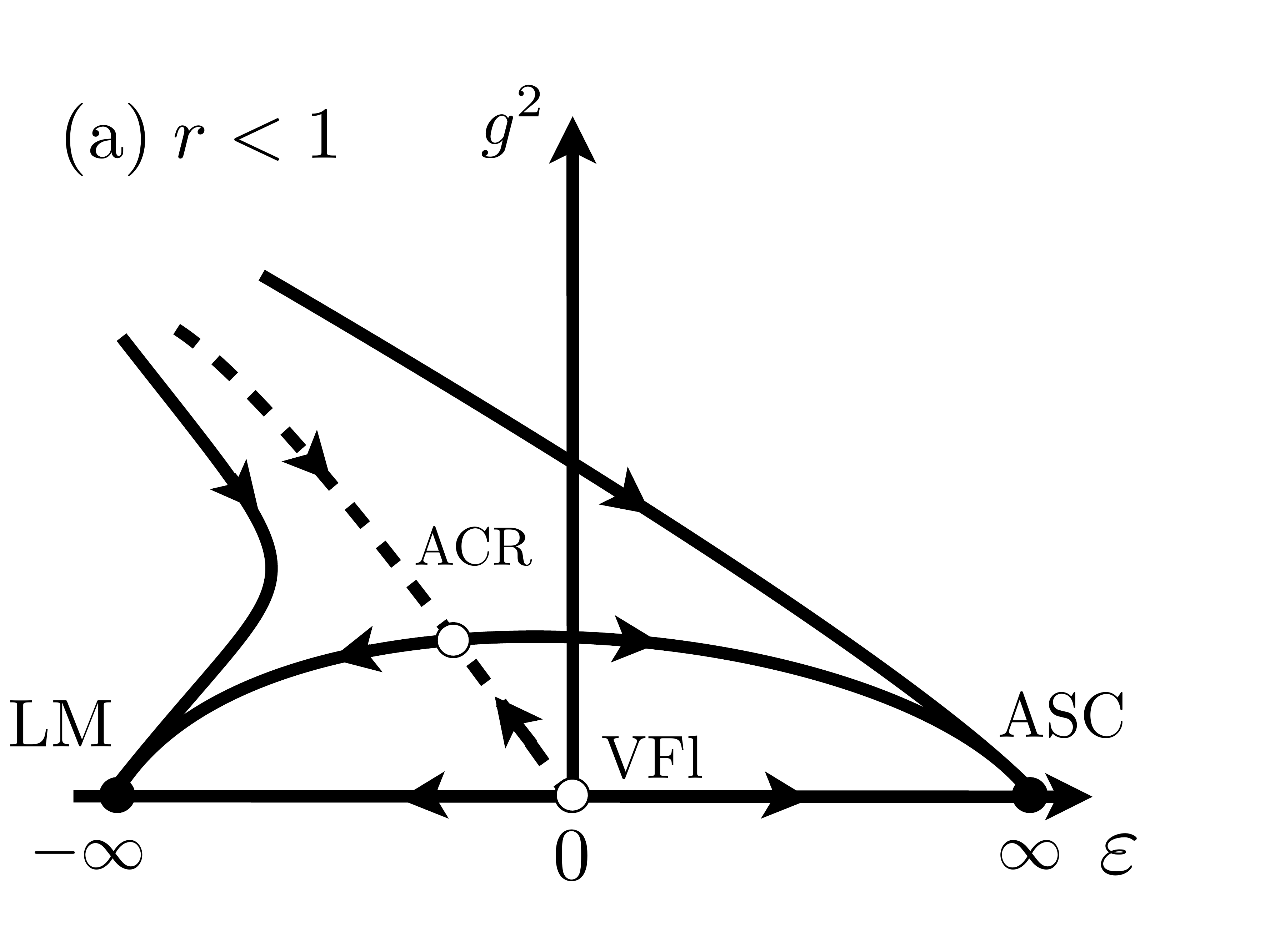}\hspace*{5mm} %
\includegraphics[width=0.41\textwidth]{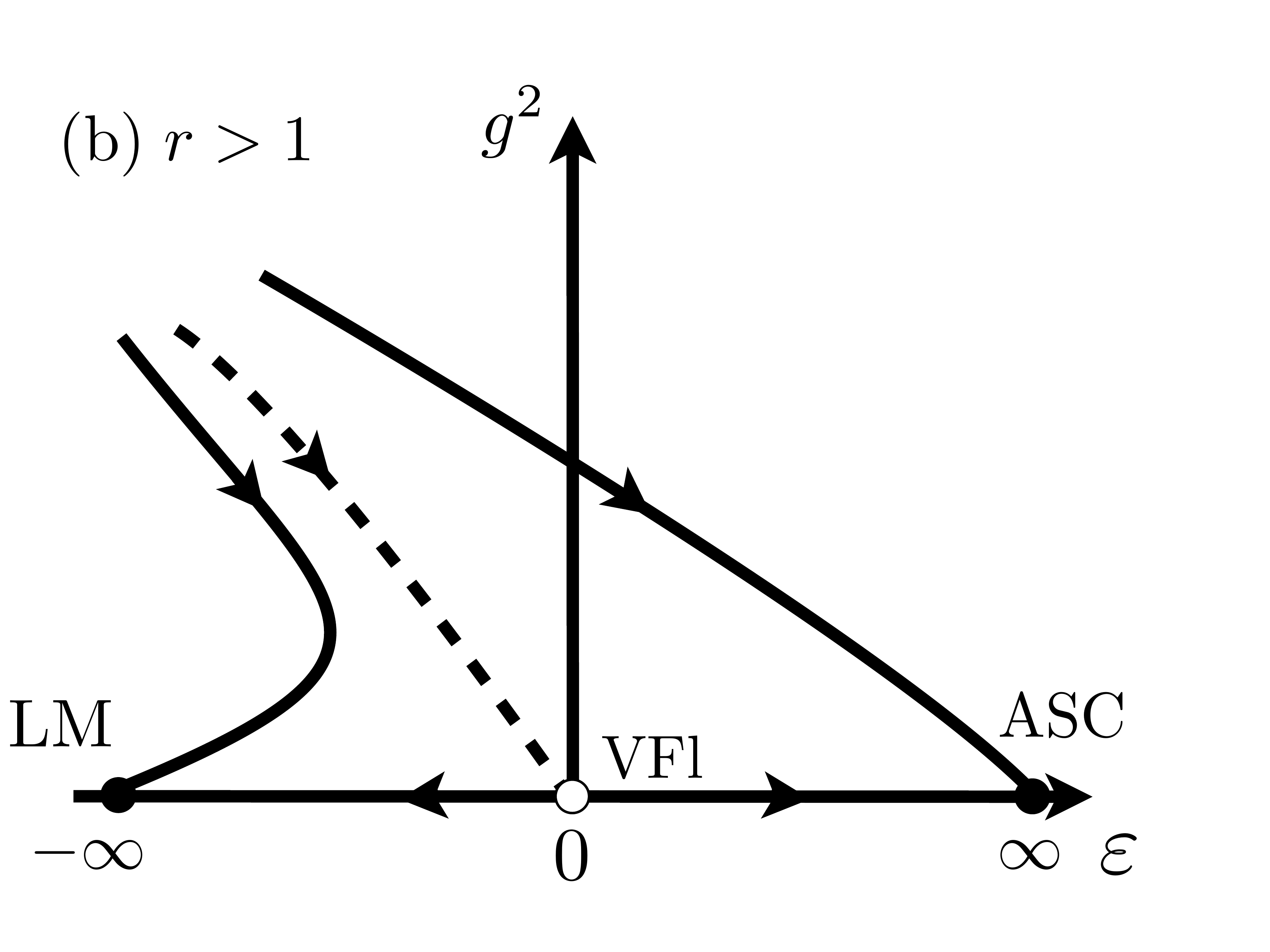}
\caption{ RG flow for the asymmetric Anderson model \protect\cite{VF04,FV04}
in the $\protect\epsilon$--$g^2$ plane, obtained from Eq.~\eqref{eq:asyrg}
(a) $r<1$: The critical fixed point ACR separates the flow towards LM from
that to ASC. (b) $r\geq1$: ASC merges with VFl as $r\to 1^-$, which
describes a level crossing with perturbative corrections. The behaviour near
$r=1$ is similar to that of the Landau-Ginzburg model near $d=4$, with VFl
and ASC corresponding to the Gaussian and Wilson-Fisher fixed points,
respectively. Reproduced from Ref.~\cite{fvrop}.}
\label{flowinfu}
\end{figure}

A perturbative expansion is now possible in $g_0$ around VFl. Power counting yields the scaling dimension of the renormalized hybridization $\dim[g] = \tilde{r}=\frac{1-r}{2}$. The one-loop flow equations for $g$ and the renormalized mass $\varepsilon$ read
\begin{eqnarray}  \label{eq:asyrg}
\frac{d g}{d \ln D}&=&-\tilde{r}g+\frac{3}{2}g^3  \notag \\
\frac{d \varepsilon}{d \ln D}&=& -\varepsilon-g^2+3g^2\varepsilon \;,
\end{eqnarray}
results to two-loop order can be found in Ref.~\cite{FV04}. The RG flow is shown in Fig.~\ref{flowinfu} -- this flow has strong similarity to that of the standard Landau-Ginzburg model. The fact that $g$ is relevant for $r<1$ and irrelevant for $r>1$ allows us to identify $r=1$ as an upper critical dimension of the pseudogap Kondo problem, akin to $d=4$ in the Landau-Ginzburg theory. For $r<1$ a non-trivial fixed point (ACR) emerges at ${g^*}^2=\frac{2}{3}\tilde{r}$ and $\varepsilon^*=-\frac{2}{3}\tilde{r}$, Fig.~\ref{flowinfu}a, similar to the celebrated Wilson-Fisher fixed point. Critical properties, evaluated in a double expansion in $\tilde{r}$ and $g$, again agree well with NRG results \cite{FV04}. In contrast, for $r\geq1$ in Fig.~\ref{flowinfu}b, we have ``Gaussian'' behaviour controlled by the VFl fixed point, which here corresponds to a simple level crossing with corrections captured by plain perturbation theory in $g_0$. In the case $r=1$, relevant to charge-neutral graphene, this perturbation theory is logarithmically divergent at criticality and needs to be resummed, as is
standard at the upper critical dimension.

\subsection{Critical quasiparticles in the pseudogap Kondo model}

After having formally analyzed the critical field theories, we may now interpret the results. We start by discussing the single-particle electronic spectra, i.e., the question of critical quasiparticles. For fermionic impurity models, the relevant quantity is the conduction-electron T matrix; in the Anderson model \eqref{eq:AIM} this is related to the $d$-electron Green's function according to $T = v^2 G_d$.

In the stable LM and ACR phases of the pseudogap Kondo model, the T-matrix spectral density inherits the behavior of the bath, ${\rm Im} T \propto |\w|^r$ \cite{GBI,FV04}. In contrast, at criticality and for $r<1$ the T matrix develops singular behavior, whose form is protected by the diagrammatic structure of the T matrix (equivalent to a Ward identity \cite{rg_bfk}) for both SCR and ACR fixed points: ${\rm Im} T \propto |\w|^{-r}$ \cite{FV04,kv04}. For $r=1$ logarithmic corrections occur, leading to ${\rm Im} T \propto 1/|\w \ln \w|^2$, whereas a free-particle $\delta(\w)$ peak is restored for $r>1$ \cite{VF04}.
We note that ${\rm Im} T \propto |\w|^{-r}$ is also realized in the SSC phase existing for $r<1/2$.

The interpretation of the critical behavior requires to distinguish the SCR and ACR fixed points. For SCR, both critical theories suggest that the criticality is exclusively carried by the spin degree of freedom of the impurity: The residual entropy is below $\ln 2$, and the Anderson-model RG flow is towards small charge fluctuations \cite{FV04}.

This is entirely different for ACR. Here, the residual entropy can exceed $\ln 2$, and for $r$ close to or larger than unity the RG flow is towards large charge flutuations \cite{VF04}. In particular, the ``Gaussian'' fixed point [i.e., the level-crossing description \eqref{aiminfu}] shows that spin and charge fluctuations are strongly coupled and equally important at criticality \cite{VF04,pixley12}. Indeed, the criticality is carried by both fermionic and bosonic degrees of freedom, $|\sigma\rangle$ and $|s\rangle$. The diagrammatic calculation shows that the spectrum of the physical $d$ electrons arises from a convolution of critical bosons and fermions, which also explains how the exponent locking, ${\rm Im} T \propto |\w|^{-r}$, is compatible with non-trivially varying exponents for other observables.


\subsection{Pseudogap Kondo physics: Applications}

\subsubsection{Kondo effect in graphene}
\label{sec:dop}

Graphene is the ideal host for the pseudogap Kondo problem: the local low-energy DOS follows Eq.~\eqref{pgdos} with $r=1$ at charge neutrality, and second-neighbor hopping $t^{\prime }\neq 0$ breaks particle--hole symmetry already on the level of the band structure. However, graphene is often doped, formally $\mu\neq 0$, due to the presence of a substrate or gates. Then, the DOS at the Fermi level is finite, and consequently a magnetic impurity described by the Kondo model \eqref{eq:Kondomodel} will be screened in the low-temperature limit for any value of the Kondo coupling $J_0$ \cite{baskaran07,cornaglia09,epl}.

On general grounds, one expects that the presence of the quantum phase transition at charge neutrality also influences the behavior at finite $\mu$. This has been studied in some detail in Ref.~\cite{epl} using a combination of analytical and numerical renormalization-group techniques, and we summarize the most interesting aspects. Near the $\mu=0$ phase transition, there exists a quantum critical regime in a plane spanned by the Kondo coupling and the doping. In this regime, the Kondo temperature is expected to follow the scaling prediction $\TK = \kappa_\pm |\mu|$ for $\mu\gtrless0$. While this applies for $r<1$, there are violations of scaling at the upper critical dimension $r=1$, such that eventually $\TK = \kappa_- |\mu|$ continues to hold for $\mu<0$, while for $\mu>0$ logarithmic corrections and Kondo logarithms conspire such that $\TK \propto |\mu|^x$ where $x\approx 2.6$ is a \textit{universal} exponent. Ref.~\cite{epl} has also given concrete predictions for the doping and parameter dependence of $\TK$ away from criticality.

Unfortunately, a clear-cut realization of Kondo physics in graphene is still missing. Existing experiments have employed Co adatoms placed on top of a graphene sheet \cite{crommie11} as well as vacancies which are believed to create a localized magnetic moment \cite{fuhrer11,nair12}. In the former case, individual Co atoms were investigated using STM techniques which indeed offer the unique opportunity to directly observe the critical quasiparticle spectra discussed above. However, the spectral features detected in Ref.~\cite{crommie11} mainly reflected charging effects and vibrational excitations. For the latter case of vacancies, signatures in low-temperature transport experiments have been interpreted in terms of Kondo physics \cite{fuhrer11}, but alternative explanations of the data have been put forward as well \cite{fuhrer_cmt}.

\subsubsection{Magnetic impurities on the surface of topological insulators}

The physics of magnetic moments coupled to boundary states of topological insulators has been discussed in numerous publications over the last years.
For 2D topological insulators, the boundaries feature 1D chiral edge channels which are expected to show Luttinger-liquid behavior. The corresponding Kondo problem has been argued to display both one-channel and two-channel Kondo phases \cite{law10}, with a possible quantum phase transition between the two \cite{chung14}.

In contrast, the surface states of strong 3D topological insulators admit a low-energy description in terms of a 2d Dirac equation. The resulting electronic properties are therefore similar to that of graphene, with a few important differences: (i) there is a single Dirac cone (or, more generally, an odd number) per surface, and (ii) the role of the pseudospin (or sublattice) in the graphene case is taken by the physical spin, such that TI surface states display spin-momentum locking, and there is no additional spin degeneracy.
The physics of Kondo impurities in this setting has been analyzed theoretically in a number of papers recently \cite{feng10,zitko10,akm:qpiti}. The main conclusion is that, despite the non-trivial topological structure of the TI surface states, the corresponding local Kondo problem for a spin $S=1/2$ impurity can be mapped onto the standard pseudogap Kondo model of Sec.~\ref{sec:pgmodel}. Consequently, the physics is similar to Kondo screening in graphene, and concrete predictions for tunneling spectra and quasiparticle interference signals have been made \cite{akm:qpiti}. To date, experimental investigations of isolated magnetic moments on TI surfaces at low temperatures are lacking.

\subsubsection{Quadratic band touching in 3D}

An interesting situation arises for quadratic band touching points\footnote{Quadratic band touching in 2D is realized, e.g., in bilayer graphene.} in 3D, as have been discussed for certain semiconductors such as HgTe \cite{tsidi97} and also in the context in pyrochlore iridates \cite{moon13}. Near a quadratic band touching point, the electronic DOS obeys $\rho(\w)\propto|\w|^r$ with $r=1/2$. Hence, one can expect that magnetic impurities in such a host realize a $r=1/2$ pseudogap Kondo model (if the host electrons are assumed to be non-interacting). This Kondo model by itself features a quantum phase transition with strong scaling behavior and critical quasiparticles (provided that particle--hole symmetry is broken). It will be highly interesting to discuss concrete settings for this physics to emerge.

\subsubsection{Weyl and Dirac semimetals in 3D}

Recently, some attention has been devoted to condensed-matter phases which feature linearly dispersing electrons in three space dimensions. This applies to Weyl semimetals \cite{volovik} which may emerge, e.g., in certain pyrochlore iridates at intermediate correlation strength \cite{wan11}, and to Dirac semimetals for which a number of materials realizations have been proposed \cite{young12,wang13}.
For Dirac electrons in 3D, the DOS follows $\rho(\w)\propto|\w|^r$ with $r=2$ if the chemical potential coincides with the energy of the Dirac point.

As with graphene, this implies the existence of a quantum phase transition for Kondo impurities, i.e., screening will only occur for sufficiently large coupling and large particle--hole asymmetry. The corresponding quantum phase transition is, however, now above the upper critical dimension, i.e., it is a level crossing with perturbative corrections \cite{VF04}. In particular, the residual local moment will jump across the transition, and the spectral function will display a $\delta$ peak with subleading corrections, i.e., critical quasiparticles will {\em not} occur.

\subsubsection{Kondo effect in spin liquids}

A somewhat exotic realization of the Kondo problem arises in the context of magnetic impurities embedded in quantum magnets. If the host magnet is in a quantum spin liquid state, the host excitations are typically spin-1/2 spinons coupled to a U(1) and Z$_2$ gauge field. The spinons couple to the impurity spin with a Kondo-like (i.e. four-point) interaction. The physics depends on the nature of the spinons, and a few cases have been discussed in the literature.

Linearly dispersing bosonic spinons yield a rich phase diagram, with a variety of possible $T=0$ phases, including the possibility of full Kondo screening, and quantum phase transitions \cite{FFV06}.
In contrast, fermionic spinons lead to physics similar to standard Kondo expectations: In
the presence of a spinon Fermi surface, the impurity spin gets always screened at low $T$
\cite{ribeiro12}. In the case of 2D Dirac spinons of an algebraic spin liquid, a quantum
phase transition not unlike that of the pseudogap Kondo problem emerges \cite{cassa,kim08,dhochak10}. However, it should be noted that the influence of gauge fields beyond perturbation theory has been neglected in the published treatments.


\section{Critical quasiparticle theory for metallic QPT}
\label{sec:pw}

The conventional theory of continuous quantum phase transitions is phrased in terms of spatial and temporal fluctuations of an order parameter leading to a Ginzburg-Landau-Wilson (GLW) $\phi^{4}$ field theory of the bosonic order parameter field. For metals this description has been pioneered by Hertz \cite{hertz}, with refinements by Millis \cite{millis} and related work by Moriya \cite{moriya}; we will below refer to this as HM theory. The order parameter field acts in $\deff=d+z$\ dimensions, with $d$ the spatial dimension and $z$ the dynamical exponent. In case that $\deff>4$ , which is the upper critical dimension of $\phi^{4}$ field theory, the theory is Gaussian, i.e. the interaction of fluctuations scales to zero at low energy under renormalization-group (RG) transformations.

How do these predictions compare with observation? Quantum phase transitions have been studied experimentally for the last several decades. In particular metallic heavy fermion compounds show a broad variety of QPTs, mostly associated with some kind of antiferromagnetic order or also with superconductivity \cite{lrvw07}. It was found that for some of the most intensely studied compounds, like CeCu$_{6-x}$Au$_{x}$ (two-dimensional antiferromagnetic (AFM) fluctuations, $\deff=4$) and YbRh$_{2}$Si$_{2}$ (three-dimensional AFM fluctuations, $\deff=5$) HM theory does not seem to apply.
One possible reason is that in these compounds one observes an apparently diverging quasiparticle effective mass $m^{\ast }$, as a function of temperature $T$ and excitation energy $\omega$ , indicating critical behavior of the quasiparticle system. In such cases it is clear that a theory of critical phenomena should account for both, critical behavior of the bosonic and
fermionic excitations. In the derivation of the GLW field theory for metallic systems fermions are integrated out, {\em assuming} they are non-critical.
Such weak-coupling-based RG theories of antiferromagnetic fluctuations in metals have
failed to predict critical properties of the kind observed in the
above-mentioned compounds \cite{lrvw07}.

One interesting option, suggested by experiment and available theory, is that the observed behavior is emerging at strong coupling. In the following a semiphenomenological theory of critical quasiparticles interacting with AFM fluctuations  \cite{wa11,aw12,asw14} will be described, resulting in a self-consistent relation for
the quasiparticle effective mass. The latter allows for two types of solutions, depending on the initial condition at high energy, first the usual
weak-coupling solution of HM theory, and secondly a new solution at strong
coupling, involving critical quasiparticles with power law divergent effective mass.
The crucial element of this theory is a Ward identity, relating the critical
quasiparticle properties to the spin--quasiparticle interaction vertex, $\lambda _{Q}\propto
m^{\ast }/m$ , at finite momentum transfer $\mathbf{q\approx Q}$, close to
the momentum corresponding to the ordering wave vector $\mathbf{Q}$.

\subsection{Critical quasiparticles}

Fermionic quasiparticle excitations are defined as poles of the
single-particle Green's function $G$

\begin{equation}
G(\mathbf{k},\omega )=\frac{1}{\omega -\epsilon _{\mathbf{k}}-\Sigma (%
\mathbf{k},\omega )}.
\end{equation}%
As we will see below the self-energy $\Sigma (\mathbf{k},\omega )$ depends only weakly on momentum $\mathbf{k}$ and we will neglect this dependence, $\Sigma (\mathbf{k},\omega )\approx \Sigma (\omega)$. Expanding at small $\omega $ one may define the quasiparticle Green's function $G^{qp}(\mathbf{k},\omega )$
\begin{equation}
G(\mathbf{k},\omega )\approx \frac{Z}{\omega -E_{\mathbf{k}}+i\Gamma }%
=ZG^{qp}(\mathbf{k},\omega ),
\end{equation}%
where the quasiparticle weight factor is defined as
\begin{equation}
Z=[1-\partial \func{Re}\Sigma /\partial \omega ]^{-1}=\frac{m}{m^{\ast
}(\omega )},
\end{equation}%
$E_{\mathbf{k}}=Z[\epsilon _{\mathbf{k}}+\Sigma (0)]$ is the renormalized
quasiparticle energy and $\Gamma =Z\func{Im}\Sigma (\omega )$ is the quasiparticle decay rate. We
extend the usual Landau quasiparticle picture into the non-Fermi liquid
regime by allowing $Z$ to be energy dependent, $Z=Z(\omega )$\ .
Quasiparticles are well-defined as long as $Z(\omega )>0$ and $\Gamma
<|\omega |$ , which is the case for a class of power law forms of the
self-energy, $\Sigma \propto \omega ^{1-\eta }$ , as long as $\eta >1/2$ .
To fully appreciate this statement one should observe that (1) the
analyticity of $\Sigma (\omega )$ in the upper half plane leads to the
relation $\func{Im}\Sigma (\omega )\propto \func{Re}\Sigma (\omega )$ if $%
\eta >0$ , (2) $\Gamma =c_{\Gamma }|\omega | $ for $0<\eta <1$ , although $\func{Im}%
\Sigma (\omega )\propto |\omega |^{1-\eta }$, where  $c_{\Gamma }<1$
if $\eta >1/2$. .

\subsection{Critical spin and energy fluctuations}

The spectrum of antiferromagnetic spin fluctuations is given by \cite{asw14}
\begin{equation}
\func{Im}\chi (\mathbf{q},\omega )=\frac{N_{0}\omega \lambda _{Q}^{2}/\gamma
}{[r+\xi _{0}^{2}(\mathbf{q-Q})^{2}]+(\omega \lambda _{Q}^{2}/\gamma )^{2}}
\label{eq:sfl-spectrum}
\end{equation}%
where $N_{0}$ is the quasiparticle density of states at the Fermi level, $\gamma
\approx v_{F}Q$ is a reference energy, and $\xi _{0}\approx 1/k_{F}$ is the
microscopic spin correlation length with $v_{F},k_{F}$ the Fermi velocity
and wave vector. The control parameter $r$ vanishes at the quantum critical
point (QCP) as $r\propto (B-B_{c})^{2\nu }$ , where $B$ is the field tuning
the QCP (magnetic field, pressure, chemical composition) $B_{c}$ marks the
critical point and $\nu $ is the correlation length exponent.\ The Landau
damping term $\omega \lambda _{Q}^{2}/\gamma $ is renormalized by two
factors of vertex functions $\lambda _{Q}\propto m^{\ast }/m$ , appearing at
both ends of a quasiparticle bubble diagram.

\subsubsection{Vertex at large momentum}

The vertex correction $\lambda _{Q}$\ follows from the Ward identity
connected with spin conservation \cite{vw14}:
\begin{equation}
\int \frac{d^{d}p^{\prime }d\omega ^{\prime }}{(2\pi )^{d+1}}[\Omega -\Delta
\epsilon _{\mathbf{p}^{\prime }}(\mathbf{Q})]\Lambda (p,p^{\prime
};Q)=\Omega -\Delta \epsilon _{\mathbf{p}}(\mathbf{Q})-\Sigma (p+Q/2)+\Sigma
(p-Q/2)
\end{equation}%
where $\Lambda (p,p^{\prime };Q)$ is the spin vertex function, $\Delta
\epsilon _{\mathbf{p}}(\mathbf{Q})=\epsilon _{\mathbf{p+Q/2}}-\epsilon _{%
\mathbf{p-Q/2}}$ . A singular behavior of the self-energy $\Sigma (p)\propto
\omega ^{1-\eta }$ , $\eta <1$ , implies $\lambda _{Q}=\lim_{\Omega
\rightarrow 0}$\ $\int_{p^{\prime }}\Lambda (p,p^{\prime };Q)\propto \omega
^{-\eta }\propto m^{\ast }(\omega )/m$.
The spectral weight given by \eqref{eq:sfl-spectrum} is peaked in momentum space at $\mathbf{q=Q}$,
implying that quasiparticles scattering off AFM spin fluctuations change their momentum
by approximately $\mathbf{Q}$ , which means that if the initial quasiparticle state was
on the Fermi surface, the final state will be ``off-shell'', far from the
Fermi surface, except for a small manifold of states on the Fermi surface,
at so-called ``hot spots''.

There are at least two ways by which the critical scattering of quasiparticles may be distributed all over the Fermi surface. The first one is by means of impurity scattering, which crudely speaking eliminates the constraints imposed by momentum conservation. This avenue has been explored in Refs.~\cite{wa11,aw12}, where power-law critical behavior of the thermodynamic and transport properties has been derived, on the basis of a self-consistent
equation for the quasiparticle effective mass. While the results for the critical
exponents of that theory appeared to be in excellent agreement with
experiment, the calculated prefactors of power laws were necessarily found
to depend on impurity concentration, in contradiction with experiment.

\subsubsection{Critical behavior of tuning parameter}

The critical behavior of the tuning parameter $r$\ on the tuning field $B$ (magnetic field, pressure, chemical composition), expressed in terms of the bare tuning parameter $r_{0}=(B-B_{c})/B_{c}$ , is generated by the critical behavior of the dressed bubble diagram $\Pi (\mathbf{q},\omega ;B)$ in the static limit $\omega =0$. The critical dependence of $\Pi $ on $B$ can be found by considering the response of $\Pi $ to  an infinitesimal change of
the tuning field, $\partial \Pi /\partial B$, at finite field $B$.
The coupling vertex to the field is renormalized by a factor $1/Z$, such that $\partial \Pi /\partial B\propto 1/Z$. Using the scaling behavior of $Z$ with energy $\omega $\ as will be calculated below, $Z(\omega )\propto \omega ^{\eta }$, and the general dynamical scaling relation $\omega \propto \xi
^{-z}$, where $\xi $ is the spin correlation length, as well as the
definition of the correlation length exponent $\xi \propto r_{0}^{-\nu }$ ,
one finds $\partial \Pi /\partial B\propto r_{0}^{-\eta z\nu }.$ Integrating
with respect to $B$ the scale-dependent contribution to $\Pi $ and hence to the
control parameter $r$ is found as $r\propto r_{0}^{1-\eta z\nu }$ . On the
other hand one has $\xi ^{-2}\propto r\propto r_{0}^{2\nu }$ . Equating the
two expressions for $r$ one finds $1-\eta z\nu =2\nu $ from which follows $%
\nu =1/(2+\eta z)$.

\subsubsection{Energy fluctuations}

A second way by which quasiparticles may become critical over the whole Fermi surface
is by scattering off-critical energy fluctuations. The spin exchange energy
operator is proportional to the scalar product of two nearest neighbor spin
operators. The simultaneous propagation of two spin propagators with
opposite momenta may therefore be viewed as energy propagation. The
corresponding spectral function is given by \cite{asw14}
\begin{equation}
\mathrm{Im}\chi _{E}(\mathbf{q},\omega )= \sum_{\mathbf{q}_{1},\omega
_{1}}G_{k+q_{1}}G_{k+q_{1}-\kappa }\mathrm{Im}\chi (\mathbf{q}_{1},\omega
_{1}) \mathrm{Im}\chi (\mathbf{q}_{1}-\mathbf{q},\omega _{1}-\omega
)[b(\omega _{1}-\omega )-b(\omega _{1})],  \notag
\end{equation}%
where $b(\omega )$ is the Bose function. The Green's functions $%
G_{k+q_{1}},G_{k+q_{1}-q}$ are off-shell for most values of the momentum $%
\mathbf{q}_{1}$\ and each may be replaced by $1/\epsilon _{F}$. Performing
the momentum integration by Fourier transform, we get
\begin{equation}
\mathrm{Im}\chi _{E}(\mathbf{q},\omega )\approx  N_{0}^{3}\lambda
_{Q}^{-2}(|\omega |\lambda _{Q}^{2}/\gamma )^{d-1/2} \frac{1}{[|\omega |\lambda _{Q}^{2}/\gamma +q^{2}\xi
_{0}^{2}+r]^{(d+1)/2}}\,.
\label{eq:enfluct}
\end{equation}
The possible relevance of these energy fluctuations near a QCP has been
noted first by Hartnoll {\em et al.} \cite{hart11}, in the context of a
calculation of the dynamical conductivity in the perturbative RG regime.

\subsubsection{Dynamical scaling}

The spectra of spin or energy fluctuations at the critical point $r=0$, as
given by \eqref{eq:sfl-spectrum} and \eqref{eq:enfluct}, have scaling properties
with regard to scaling of energy and wavevector
\begin{equation}
\omega \propto q^{z}\,,~~~~~~z=2/(1-2\eta )\,.
\end{equation}%
It follows that the (bosonic) correlation length in the critical regime
diverges as a function of temperature as $\xi \propto T^{-1/z}$.

Defining a correlation length
\begin{equation}
\xi (\omega ;r)=[\omega \lambda _{Q}^{2}/\gamma +r]^{-1/2},
\label{eq:corrlength}
\end{equation}%
we find that
\begin{eqnarray}
\mathrm{Im}\chi (\mathbf{q},\omega )& \propto & \frac{\omega \lambda _{Q}^{2}}{%
[\xi ^{-2}(\omega ;r)+(\mathbf{q-Q})^{2}]^{2}}\,,  \notag \\
\mathrm{Im}\chi _{E}(\mathbf{q},\omega ) &\propto & \frac{\lambda
_{Q}^{-2}(\omega \lambda _{Q}^{2})^{d-1/2}}{[\xi ^{-2}(\omega
;r)+q^{2}]^{(d+1)/2}}\,.
\end{eqnarray}

\subsection{Critical effective mass}

The electron self-energy $\Sigma $ may now be calculated in leading (one-loop) order. The imaginary part of $\Sigma $ is obtained as
\begin{equation}
\mathrm{Im}~\mathrm{\Sigma (k,\omega )} \mathrm{=-\lambda
_{E}^{2}\int \frac{d\nu }{\pi }\sum_{\mathbf{q}}\mathrm{Im}G(\mathbf{k}+%
\mathbf{q},\omega +\nu )} \mathrm{Im}\chi _{E}(\mathbf{q},\nu )[b(\nu )+f(\omega -\nu )].
\notag
\end{equation}
The interaction vertex $\lambda _{E}=\lambda _{Q}^{2}\lambda _{v}$, where $\lambda _{v}$ is $\propto m^{\ast }/m$, as it arises through a Ward identity
connected to energy conservation, and $\lambda _{Q}\propto m^{\ast }/m$,
as discussed above. The Fermi and Bose functions $f(\omega ),b(\omega )$
confine the $\nu $-integration at low $T$ to the interval $[0,\omega ]$. The
momentum integration involves integration over the angle $\theta $ enclosed
by the momenta $\mathbf{k,q}$.
Using $\mathrm{\mathrm{Im}G(\mathbf{k+q},\omega )\approx \pi Z\delta (\omega -Z\epsilon }_{\mathbf{k+q}})$ and $\mathrm{\epsilon }_{\mathbf{k+q}}\approx \mathrm{\epsilon }_{\mathbf{k}}+\mathbf{v}_{\mathbf{k}}\cdot \mathbf{q}$ , where $\mathbf{v}_{\mathbf{k}}\cdot \mathbf{q=}v_{F}q\cos \theta $ one finds
\begin{equation}
\int d\cos \theta \, \mathrm{Im}G(\mathbf{k},\omega)\approx \frac{1}{v_{F}q}\,\Theta \left(v_{F}q-|\frac{\omega +\nu}{Z}-\epsilon_{\mathbf{k}}|\right)\,.
\end{equation}%
Here $\Theta (x)$ is the step function. Performing the momentum
and frequency integrations, using $\mathrm{Im}\chi _{E}(\mathbf{q},\nu )$ as
given by \eqref{eq:enfluct} one finds for quasiparticles on the Fermi surface ($\mathrm{%
\epsilon }_{\mathbf{k}}=0$)
\begin{equation}
\mathrm{Im}~\mathrm{\Sigma (k,\omega )}\approx \gamma \lambda
_{Q}^{2}(\omega \lambda _{Q}^{2}/\gamma )^{d+1/2}\xi ^{2}(\omega ;r)
\label{eq:sigmascale}
\end{equation}%
where $\xi (\omega ;r)$ has been defined by \eqref{eq:corrlength}, observing that the step function in \eqref{eq:enfluct} is preempted by the condition $q>\xi ^{-1}$, which defines the dominant regime of the $q$ integral.

In the critical regime ($r=0$)\ we may now substitute the power law $\lambda_{Q}(\omega )\propto m^{\ast }(\omega )/m\propto |\omega |^{-\eta }$ into \eqref{eq:sigmascale} to obtain the scale-dependent contribution to $\mathrm{Im}$ $\mathrm{\Sigma }$ as
\begin{equation}
\mathrm{Im}~\mathrm{\Sigma (k,\omega )}\propto |\omega |^{d-1/2-\eta
(2d+1)}\,.
\end{equation}

The scale-dependent contribution to $\mathrm{Re}\Sigma (\omega )$ follows
from analyticity as $\mathrm{Re}\Sigma (\omega )\propto (\lambda
_{Q})^{2d+1}(\omega /\gamma )^{d-1/2}\propto $ $[m^{\ast }(\omega
)/m]^{2d+1}(\omega /\gamma )^{d-1/2}$.
Substituting this into the definition of the effective mass one finds the
self-consistency relation for $m^{\ast }(\omega )$ \cite{asw14}
\begin{equation}
\frac{m^{\ast }(\omega )}{m}=1+\left[\frac{m^{\ast }(\omega )}{m}\right]^{2d+1}\left(\frac{%
\omega }{\gamma }\right)^{d-3/2}\,.  \label{eq:scmstar}
\end{equation}%
This equation, for dimensions $d>3/2$, has two types of solution, depending on the initial condition at high energy, at the upper end of the scaling regime. The first one is the usual H-M-model solution, for which the second term on the r.h.s. is always less than unity, implying that $m^{\ast}(\omega )/m\approx 1$ and the quasiparticle mass is not critical. If, however, the effective mass is already enhanced by some additional fluctuation contribution, such that the scale dependent term $\propto (\frac{\omega }{\gamma })^{d-3/2}$ is already larger than unity, a new and self-consistent power law solution emerges,
\begin{eqnarray}
\frac{m^{\ast }(\omega )}{m} &\approx &\left(\frac{\omega }{\gamma }\right)^{1/2-3/4d}\propto \omega ^{-\eta }\,,  \notag \\
\eta  &=&\frac{1}{2}-\frac{3}{4d}=\QATOPD\{ . {1/4~~~d=3}{1/8%
~~~d=2}~.
\end{eqnarray}%
The critical exponents follow as
\begin{eqnarray}
z &=&\frac{2}{1-2\eta }=\frac{4d}{3}\,,  \notag \\
\nu  &=&\frac{1}{2+\eta z}=\frac{3}{3+2d}\,.
\end{eqnarray}

\subsubsection{Renormalization-group equation}

The self-consistent equation \eqref{eq:scmstar} for the effective mass describes scaling of the coupling constant $Y=m^{\ast }/m-1$ as a function of the scaling variable $\Lambda =-\ln (\omega/\gamma )$. The solution of \eqref{eq:scmstar} may be thought of as a result of solving a renormalization group (RG) equation for $Y(\Lambda )$. The corresponding RG equation is easily found as
\begin{equation}
\frac{dY}{d\Lambda }=\beta (Y)=\eta \frac{Y(Y+1)}{Y-Y_{u}}
\end{equation}%
where $Y_{u}=1/2d$. The $\beta $-function is seen to be negative for $Y<Y_{u}$, so that for any initial condition at $\Lambda _{0}$, $Y(\Lambda_{0})<Y_{u}$ the RG flow is directed away from $Y_{u}$, towards the first stable fixed point at $Y=0$. Beyond the unstable fixed point at $Y_{u}$ the $\beta $-function is positive, and the RG flow is directed away from $Y_{u}$ to the second stable fixed point at $Y\rightarrow \infty$. Since $\lim_{Y\rightarrow \infty }\beta (Y)=\eta Y$, the power-law exponent describing the divergence of $Y$ is given by $\eta $, $Y\propto \omega^{-\eta}$, in agreement with the direct solution of Eq.~\eqref{eq:scmstar}.

\subsection{Thermodynamic scaling}

The free energy of the system is dominated by the critical contribution of
the quasiparticles and may be calculated employing the expression for the
entropy density $S/V$ in terms of the self-energy
\begin{equation}
\frac{S}{V}\mathrm{=N(0)\int \frac{d\omega }{2\pi }}\frac{\omega (\omega -%
\func{Re}\Sigma (\omega ))}{T^{2}\cosh ^{2}\frac{\omega }{2T}}.
\label{eq:entropy}
\end{equation}%
For a critical discussion of \eqref{eq:entropy}\ see \cite{chub05}.
Substituting the self-energy as determined by \eqref{eq:sigmascale} using
again the analytical properties of $\Sigma (\omega )$ in the complex
frequency plane one finds $S/V$ as a function of the scaling variables
temperature $T\xi ^{z}$ and control parameter $r_{0}\xi ^{1/\nu }$.
Integrating with respect to temperature one finally gets the free energy
density
\begin{equation}
\frac{F(T,B)}{V}\mathrm{=\xi }_{f}^{-(d_{f}+z_{f})}\mathrm{\Phi }_{f}(T\xi
_{f}^{z_{f}},r_{0}\xi _{f}^{1/\nu _{f} }).  \label{eq:free_energy}
\end{equation}%
Here $\xi _{f}$ is the correlation length and $z_{f},\nu _{f}$ are the critical exponents of the fermionic degrees of freedom.  One observes that
the free energy density obeys hyperscaling. The spatial dimension of the fermions is $d_{f}=1$,
since momentum enters through the quasiparticle energy, which depends only on the distance from the Fermi surface. The dynamical exponent then follows from the quasiparticle Green's function as $z_{f}=1/(1-\eta )$ , which determines $\mathrm{\xi }_{f}$\ $\propto T^{-1/z_{f}}$. Alternatively one may represent the free energy density in terms of the bosonic correlation length $\xi \propto T^{-1/z}$, in which case the correlation volume may be expressed as $\mathrm{\xi }_{f}^{d_{f}+z_{f}}=$ $\xi ^{2d+1}$. We conclude that while the primary critical entities, the fermions, satisfy hyperscaling, the critical bosons
do not. The bosons are driven by the fermions into a critical state
intermediate between Gaussian and fully critical. We recall that in the
cases considered here the bosonic fluctuations in the case of weak coupling
to the fermions would be Gaussian. In the case of strong coupling\ we expect
that the direct boson-boson interaction still flows to zero in the RG
process.

Employing the scaling form of the free energy, Eq.~\eqref{eq:free_energy}, the critical contributions to the thermodynamic quantities, both in the critical regime (C) and in quantum disordered Fermi-liquid (FL) regime may be obtained. The specific heat coefficient $\gamma =C/T$ is found to diverge as $\gamma \propto T^{-\eta }$ and $\gamma \propto r_{0}^{-(2d-3)/(2d+3)}$ in the C and FL regimes, respectively.
The critical part of the polarization induced by the control parameter $B$ (magnetization $M$, density $\rho $, \ldots) is found as $M(T,B_{c})-M(0,B_{c})\propto -T$ \ (C) and $M(0,B)-M(0,B_{c})\propto r_{0}^{4d/(2d+3)}$ (FL). The corresponding susceptibilities $\chi =\partial M/\partial B$ receive critical contributions $\chi (T,B_{c})-\chi (0,B_{c})\propto -T^{(2d-3)/4d}$ (C) and $\chi (0,B)-\chi (0,B_{c})\propto r_{0}^{(2d-3)/(2d+3)}$ (FL).
A quantity of special interest is the Gr\"{u}neisen ratio $\Gamma _{G}=-(\partial M/\partial T)/C$ for which we find $\Gamma _{G}\propto C^{-1}\propto T^{-(2d+3)/4d}$ (C) and $\Gamma _{G}=-G_{r}/(B-B_{c})$ (FL), where $G_{r}=-(2d-3)/(2d+3)$ is a universal coefficient \cite{garst03}.

\subsection{Scaling of transport properties}

The transport properties may be accessed in the framework of the critical
quasiparticle picture by extending the relations known from Fermi liquid
theory into the critical regime. The electrical resistivity $\rho$, for
example, in FL theory is given by $\rho =m^{\ast }\Gamma /e^{2}n$, where $e$
and $n$ are the charge and the density of the carriers and $\Gamma$ is the
scattering relaxation rate. In the case that a typical momentum transfer in
a scattering process is large, either intrinsically or by mediation of
impurity scattering, one may approximate $\Gamma$ by the imaginary part of
the self-energy. Quite generally, one expects $\Gamma$, as a quantity of
dimension energy, to satisfy the scaling relation
\begin{equation}
\Gamma \mathrm{=\xi }^{-z}\mathrm{\Phi }_{\Gamma }(T\xi ^{z},r_{0}\xi
^{1/\nu })\,.
\end{equation}%
Combining this with the scaling of the effective mass, one arrives at a
scaling form for the inelastic part of the resistivity
\begin{eqnarray}
\rho (T,B)-\rho (T,B_{c}) &\propto &\frac{m^{\ast }}{m}\mathrm{\xi }^{-z}%
\mathrm{\Phi }_{\Gamma }(T\xi ^{z},r_{0}\xi ^{1/\nu }) \\
&\propto &\QATOPD\{ . {T^{(2d+3)/4d}~~~~~~~~~~~{\rm(C)}%
}{r_{0}^{-3(2d-1)/(3+2d)}T^{2}~~{\rm(FL)}} ~~.
\end{eqnarray}

\subsection{Critical quasiparticles at quantum critical points in bulk
metals: Summary}

The results presented in this section have been compared with experimental
data for two heavy fermion compounds, CeCu$_{6-x}$Au$_{x}$ \cite{hvl96} and YbRh$_{2}$Si$_{2}$ \cite{gegen02}. In both cases the specific heat measurements show an apparently diverging effective mass in the limit $T\rightarrow 0$. In the first case the critical point is located at
concentration $x=0.1$ for ambient pressure and in zero magnetic field, or
may be tuned by pressure and/or magnetic field away from $x=0.1$. In the
second case the transition is field tuned at ambient pressure and the
critical field is $B_{c}=60$~mT. Neutron scattering studies indicate
two-dimensional antiferromagnetic fluctuations in CeCu$_{6-x}$Au$_{x}$ \cite%
{schroed98}, with  ferromagnetic fluctuations superposed \cite{rosch97}.
Three-dimensional antiferromagnetic fluctuations have been identified in YbRh%
$_{2}$Si$_{2}$ at low temperatures, $T\lesssim 0.3$~K, whereas ferromagnetic
fluctuations prevail at higher $T$ \cite{broholm12}. Applying the theory
presented above assuming two-dimensional fluctuations (in the
three-dimensional metal) the specific heat coefficient is found to diverge
as $\gamma \propto T^{-1/8}$ (rather than logarithmically, as conventionally
assumed), in very good agreement with the data on CeCu$_{6-x}$Au$_{x}$  \cite%
{asw14}. The theoretical result for the resistivity is $\rho (T)-\rho
(0)\propto T^{7/8}$ (rather than linear), again in very good agreement with
the data \cite{asw14}. A crucial test of the theory is the scaling behavior
of the dynamical structure factor observed in inelastic neutron scattering
studies \cite{schroed98} , which allowed an excellent fit to a power law
scaling form with exponent of value $\alpha \approx 3/4$, which is matched
by our theory where $\alpha =1-2\eta =3/4$. Applying the theory assuming
three-dimensional fluctuations to YbRh$_{2}$Si$_{2}$ at low temperature and
magnetic field, $T\lesssim 0.3$~K, $B\lesssim 0.3$~T one again finds excellent
agreement with the available data \cite{wa11,aw12}. The specific heat
coefficient is found to diverge as $\gamma \propto T^{-1/4}$ and $\gamma
\propto (B-B_{c})^{-1/3}$ in the C- and FL- regimes, respectively. The
resistivity shows a sublinear $T$-dependence, $\rho (T)-\rho (0)\propto
T^{3/4}$ in the C-regime as seen in the data. The agreement of the scaling
contributions with data on the magnetization, the spin susceptibility, the
thermopower,  and other quantities is also excellent \cite{wa11,aw12}. Of
particular interest is the Gr\"{u}neisen ratio in the FL-regime, which is
characterized by the universal number $G_{r}$ , for which we find $G_{r}=-1/3
$, while the experimental value is $G_{r}\approx -0.3$ .

It appears that the self-consistent theory of quantum criticality \cite%
{wa11,aw12,asw14} captures essential features of the quantum critical
behavior of metals in situations where the quasiparticle effective mass diverges, but
the interaction of bosonic fluctuations is still weak. One is then in an
intermediate regime of strong-coupling character with respect to the
fermion-boson-interaction, but still in the weak-coupling regime of the
boson-boson interaction. Anticipating that the one-loop approximation of the
self-energy (employed in the above) is accurate, the theory would then be
considered as a dynamically renormalized Gaussian theory.


\section{Conclusions}
\label{sec:summ}

We have reviewed the theoretical description of two cases of quantum phase transitions where fermionic spectral functions develop power-law singularities, a situation dubbed ``critical quasiparticles''.
The first case was about boundary phase transitions in the pseudogap Kondo model: Here the critical behavior is fully understood thanks to the existence of suitable epsilon expansion schemes. Possible experimental realizations include the Kondo effect in graphene, on the surface of topological insulators, and in systems with quadratic band-touching points.
The second case was about an antiferromagnetic transition in a metal with strong coupling between fermions and spin fluctuations. Here the existing self-consistent theory is semiphenomenological, and a fully controlled derivation from a microscopic model is not yet available.

We believe that various other quantum phase transitions, in particular those which involve the onset of Mott localization, should also display critical quasiparticles. Developing effective theories for those is an important task for future work.


\acknowledgement

We thank E. Abrahams, F. B. Anders, A. Benlagra, L. Fritz, A. Mitchell, A. Rosch, I. Schneider, and J. Schmalian for collaborations and fruitful discussions on this subject.
This research was supported by the Deutsche Forschungsgemeinschaft through the Research Unit FOR 960.



\begin{thebibliography}{99}

\bibitem{ssbook} S.~Sachdev, \textit{Quantum Phase Transitions}, Cambridge
University Press, Cambridge (1999).

\bibitem{mvrop}
M. Vojta,
Rep. Prog. Phys. {\bf 66}, 2069 (2003).

\bibitem{lrvw07} H. von L\"{o}hneysen, A. Rosch, M. Vojta, and P. W\"{o}lfle,
Rev. Mod. Phys. \textbf{79}, 1015 (2007).

\bibitem{ss_rev08}
S. Sachdev,
Nat. Phys. {\bf 4}, 173 (2008).

\bibitem{steg_rev08}
P. Gegenwart, Q. Si, and F. Steglich,
Nat. Phys. {\bf 4}, 186 (2008).

\bibitem{gia_rev08}
T. Giamarchi, C. R\"uegg, and O. Tchernyshyov,
Nat. Phys. {\bf 4}, 198 (2008).

\bibitem{tv_rev06}
T. Vojta,
J. Phys. A {\bf 39}, R143 (2006).

\bibitem{imada_rev98}
M. Imada, A. Fujimori, and Y. Tokura,
Rev. Mod. Phys. {\bf 70}, 1039 (1998).

\bibitem{xu12}
C. Xu,
Int. J. Mod. Phys. B {\bf 26}, 18 (2012).

\bibitem{fiete12}
G. A. Fiete, V. Chua, M. Kargarian, R. Lundgren, A. R\"uegg, J. Wen, and V. Zyuzin,
Physica E {\bf 44}, 845 (2012)

\bibitem{mvrev} M. Vojta, Phil. Mag. \textbf{86}, 1807 (2006).

\bibitem{hewson} A.~C.~Hewson, \emph{The Kondo Problem to Heavy Fermions},
Cambridge University Press, Cambridge (1997).

\bibitem{flst2}
T. Senthil, M. Vojta, and S. Sachdev,
Phys. Rev. B \textbf{69}, 035111 (2004).

\bibitem{senthil08}
T. Senthil,
Phys. Rev. B {\bf 78}, 045109 (2008).

\bibitem{osmottrev}
M. Vojta,
J. Low Temp. Phys. {\bf 161}, 203 (2010).

\bibitem{dobro13}
J. Vucicevic, H. Terletska, D. Tanaskovic, and V. Dobrosavljevic,
Phys. Rev. B {\bf 88}, 075143 (2013).

\bibitem{metzner03}
W. Metzner, D. Rohe, and S. Andergassen,
Phys. Rev. Lett. {\bf 91}, 066402 (2003).

\bibitem{edmft} Q. Si, S. Rabello, K. Ingersent, and J.~L. Smith, Nature
(London) \textbf{413}, 804 (2001).

\bibitem{dmft} W.~Metzner and D.~Vollhardt, Phys. Rev. Lett. \textbf{62},
324 (1989); A.~Georges, G.~Kotliar, W.~Krauth and M.~J.~Rozenberg, Rev. Mod.
Phys. \textbf{68}, 13 (1996).


\bibitem{withoff} D.~Withoff and E.~Fradkin,
Phys. Rev. Lett. \textbf{64}, 1835 (1990).


\bibitem{ingersent} K.~Ingersent, Phys. Rev. B \textbf{54}, 11936 (1996).

\bibitem{bulla97}
R. Bulla, T. Pruschke, and A. C. Hewson,
J. Phys.: Condens. Matter {\bf 9}, 10463 (1997).

\bibitem{GBI} C.~Gonzalez-Buxton and K.~Ingersent, Phys. Rev. B \textbf{57},
14254 (1998).


\bibitem{si02} K.~Ingersent and Q.~Si, Phys. Rev. Lett. \textbf{89}, 076403
(2002).

\bibitem{nrg} K.~G.~Wilson, Rev. Mod. Phys. \textbf{47}, 773 (1975).

\bibitem{nrg_rev} R. Bulla, T. Costi, and T. Pruschke, Rev. Mod. Phys.
\textbf{80}, 395 (2008).

\bibitem{VF04} M. Vojta and L. Fritz, Phys. Rev. B \textbf{70}, 094502
(2004).

\bibitem{FV04} L. Fritz and M. Vojta, Phys. Rev. B \textbf{70}, 214427
(2004).

\bibitem{fvrop}
L. Fritz and M. Vojta,
Rep. Prog. Phys. \textbf{76}, 032501 (2013).

\bibitem{schneider11}
I. Schneider, L. Fritz, F. B. Anders, A. Benlagra, and M. Vojta,
Phys. Rev. B {\bf 84}, 125139 (2011).

\bibitem{cazalilla12} M. A. Cazalilla, A. Iucci, F. Guinea, and A. H. Castro
Neto, preprint arXiv:1207.3135.

\bibitem{mf13} A. K. Mitchell and L. Fritz, Phys.~Rev.~B \textbf{88}, 075104
(2013).

\bibitem{fracsp}
A. Mitchell, M. Vojta, R. Bulla, and L. Fritz,
Phys. Rev. B {\bf 88}, 195119 (2013).

\bibitem{poor} P.~W.~Anderson, J. Phys. C \textbf{3}, 2436 (1970).

\bibitem{rg_bfk} L. Zhu and Q. Si, Phys. Rev. B \textbf{66}, 024426 (2002);
G. Zarand and E. Demler, Phys. Rev. B \textbf{66}, 024427 (2002).


\bibitem{kv04} M. Kir\'{c}an and M. Vojta, Phys. Rev. B \textbf{69}, 174421
(2004).

\bibitem{pixley12}
J. H. Pixley, S. Kirchner, K. Ingersent, and Q. Si,
Phys. Rev. Lett. {\bf 109}, 086403 (2012).

\bibitem{baskaran07} 
K. Sengupta and G. Baskaran, Phys.~Rev.~B \textbf{77}, 045417 (2008).

\bibitem{cornaglia09}
P. S. Cornaglia, G. Usaj, and C. A. Balseiro,
Phys. Rev. Lett. {\bf 102}, 046801 (2009).

\bibitem{epl}
M. Vojta, L. Fritz, and R. Bulla,
EPL \textbf{90}, 27006 (2010).

\bibitem{crommie11}
V. W. Brar, R. Decker, H.-M. Solowan, Y. Wang, L. Maserati, K. T. Chan, H. Lee, C. O. Girit, A. Zettl, S. G. Louie, M. L. Cohen, and M. F. Crommie,
Nature Phys. {\bf 7}, 43 (2011).

\bibitem{fuhrer11} 
J. H. Chen, L. Li, W. G. Cullen, E. D. Williams, and M. S. Fuhrer, Nature
Phys. \textbf{7}, 535 (2011).

\bibitem{nair12}
R. R. Nair, M. Sepioni, I.-L. Tsai, O. Lehtinen, J. Keinonen, A. V. Krasheninnikov, T. Thomson, A. K. Geim, and I. V. Grigorieva, Nature Phys. {\bf 8}, 199 (2012).

\bibitem{fuhrer_cmt} H. Jobst and H. B. Weber, Nature Phys. \textbf{8}, 352
(2012); J.-H. Chen \emph{et al.} \emph{ibid.} 353 (2012).




\bibitem{law10}
K. T. Law, C. Y. Seng, P. A. Lee, and T. K. Ng,
Phys. Rev. B \textbf{81}, 041305 (2010).

\bibitem{chung14}
C.-H. Chung and S. Silotri,
New J. Phys. {\bf 17}, 013005 (2015).

\bibitem{feng10}
X.-Y. Feng, W.-Q. Chen, J.-H. Gao, Q.-H. Wang, and F. C. Zhang,
Phys. Rev. B {\bf 81}, 235411 (2010).

\bibitem{zitko10} R. Zitko, Phys. Rev. B \textbf{81}, 241414(R) (2010).

\bibitem{akm:qpiti} A.~K.~Mitchell, D.~Schuricht, M.~Vojta, and L.~Fritz,
Phys. Rev. B \textbf{87}, 075430 (2013).





\bibitem{tsidi97}
I. M. Tsidilkovski, {\em Electron Spectrum of Gapless Semiconductors},
Springer-Verlag, Berlin, 1997.

\bibitem{moon13}
E.-G. Moon, C. Xu, Y. B. Kim, and L. Balents,
Phys. Rev. Lett. {\bf 111}, 206401 (2013).

\bibitem{volovik}
G. E. Volovik, {\em The Universe in a Helium Droplet}, (Clarendon, Oxford, 2003);
Lect. Notes Phys. {\bf 718}, 31 (2007).

\bibitem{wan11}
X. Wan, A. M. Turner, A. Vishwanath, and S. Y. Savrasov,
Phys. Rev. B {\bf 83}, 205101 (2011).

\bibitem{young12}
S. M. Young, S. Zaheer, J. C. Y. Teo, C. L. Kane, and E. J. Mele,
Phys. Rev. Lett. {\bf 108}, 140405 (2012).

\bibitem{wang13}
Z. J. Wang, H.-M. Weng, Q.-S. Wu, X. Dai, and Z. Fang,
Phys. Rev. B {\bf 88}, 125427 (2013).

\bibitem{FFV06} S. Florens, L. Fritz, and M. Vojta,
Phys. Rev. Lett. \textbf{96}, 036601 (2006).

\bibitem{ribeiro12} P. Ribeiro and P. A. Lee, Phys. Rev. B \textbf{83},
235119 (2011).

\bibitem{cassa}
C. R. Cassanello and E. Fradkin,
Phys. Rev. B {\bf 53}, 15079 (1996).

\bibitem{kim08} K.-S. Kim and M. D. Kim, J. Phys. Cond. Matter \textbf{20},
125206 (2008).

\bibitem{dhochak10} K. Dhochak, R. Shankar, and V. Tripathi, Phys. Rev.
Lett. \textbf{105}, 117201 (2010).


\bibitem{hertz}
J.~A.~Hertz,
Phys. Rev. B {\bf 14}, 1165, (1976).

\bibitem{millis}
A.~J.~Millis,
Phys. Rev. B {\bf 48}, 7183, (1993).

\bibitem{moriya} T.~Moriya, {\em Spin Fluctuations
in Itinerant Electron Magnetism}, Springer-Verlag, Berlin (1985);
T.~Moriya and T.~Takimoto, J. Phys. Soc. Jpn. {\bf 64}, 960
(1995).

\bibitem{vw14} C. M. Varma and P. W\"{o}lfle, to be published.

\bibitem{wa11} P. W\"{o}lfle and E. Abrahams, Phys. Rev. B \textbf{84},
041101 (2011).

\bibitem{aw12} E. Abrahams and P. W\"{o}lfle, Proc. Nat. Acad. Sciences
\textbf{109}, 3238 (2012).

\bibitem{asw14} E. Abrahams, J. Schmalian, and P. W\"{o}lfle,
Phys. Rev. B \textbf{90}, 045105 (2014).

\bibitem{hart11} S. A. Hartnoll, D. M. Hofman, M. A. Metlitski, and S. Sachdev,
Phys. Rev. B \textbf{84}, 125115 (2011).

\bibitem{chub05} A. V. Chubukov, D. L. Maslov, S. Gangadharaya, and L. Glazman,
Phys. Rev. B \textbf{71}, 205112 (2005).

\bibitem{garst03} L. Zhu, M. Garst, A. Rosch, and Q. Si,
Phys. Rev. Lett. \textbf{91,} 066404\ (2003).

\bibitem{rosch97}
A. Rosch, A. Schr\"oder, O. Stockert, and H. von L\"ohneysen,
Phys. Rev. Lett. {\bf 79}, 159 (1997).

\bibitem{schroed98} A. Schr\"{o}der, G. Aeppli, E. Bucher, R. Ramazashvili,
and P. Coleman, Phys. Rev. Lett. \textbf{80}, 5623 (1998).

\bibitem{hvl96} H. von L\"{o}hneysen, J. Phys.: Condens. Matter \textbf{8},
9698 (1996).

\bibitem{broholm12}
C. Stock, C. Broholm, F. Demmel, J. van Duijn, J. W. Taylor, H. J. Kang, R. Hu, and C. Petrovic,
Phys. Rev. Lett. \textbf{109}, 127201 (2012).

\bibitem{gegen02} P. Gegenwart, J. Custers, C. Geibel, K. Neumaier, T. Tayama, K. Tenya, O. Trovarelli, and F. Steglich, Phys. Rev. Lett. \textbf{89}, 056402 (2002).


\end{thebibliography}
\end{document}